\documentclass[pdflatex,sn-mathphys]{sn-jnl}
\usepackage{xcolor}
\usepackage[utf8]{inputenc}
\jyear{2022}%

\begin{document}
	
	\title[ ]{Horizontal flux growth as an efficient preparation method of CeRh$_2$As$_2$ single crystals}

	\author*{\fnm{Grzegorz} \sur{Chajewski}}\email{g.chajewski@intibs.pl}
	
	\author{\fnm{Damian} \sur{Szymański}}\email{d.szymanski@intibs.pl}
	
	\author{\fnm{Dariusz} \sur{Kaczorowski}}\email{d.kaczorowski@intibs.pl}
	
	\affil{\orgname{Institute of Low Temperature and Structure Research, Polish Academy of Sciences}, \orgaddress{\street{Ok\'olna 2}, \postcode{50-422} \city{Wroc{\l}aw}, \country{Poland}}}

	
	\abstract{We report on the efficient method of obtaining CeRh$_2$As$_2$ single crystals with the use of a bismuth flux growth method in a horizontal configuration. The crystals thus obtained exhibit much sharper phase transitions and distinctly higher characteristic temperatures $T_{\rm c}$ and $T_0$, compared to previous reports.  Moreover, in our numerous attempts, we found this technique to be scalable and repeatable.}
	
	
	
	
	\maketitle
	
	\section*{Introduction}\label{sec1}
	
	The discovery of multi-phase superconductivity in CeRh$_2$As$_2$ \cite{Khim2021} turned the eyes of many solid-state researchers toward this compound and made it almost instantly one of the most highlighted materials of recent times.  In consequence, numerous papers, both experimental \cite{Khim2021, Hafner2022, Landaeta2022, Onishi2022, Kimura2021, Kibune2022, Kitagawa2022} and theoretical \cite{Schertenleib2021, Mockli2021a, Mockli2021b, Ptok2021, Cavanagh2022, Nogaki2021, Hazra2022, Nogaki2022, Machida2022}, have been published in the last few years. The most remarkable features of this material are related to the phase transition within the superconducting state, which occurs upon applying a magnetic field along the \textit{c}-axis and leads to an extraordinarily distinct violation of the Pauli-Clogston limit. Although several different phenomena including local inversion symmetry breaking \cite{Khim2021}, quadrupole density wave \cite{Khim2021, Hafner2022}, and coexisting antiferromagnetism \cite{Machida2022} have been proposed as the origin of this behavior, there is no clear experimental evidence in favor of either of them so far. One of the most essential difficulties to resolve the puzzle of the CeRh$_2$As$_2$ nature is related to the quality of samples available up to date. In all of them, a significant blurring of the anomalies associated with the phase transitions at $T_0$ and $T_{\rm c}$ is observed. This, in consequence, makes the analysis of collected experimental data more complicated. 
	
	In order to properly understand the physics governing the unusual behavior of CeRh$_2$As$_2$, addressing two basic issues seems to be crucial: 1) \textit{Which physical mechanism is responsible for the phase transition within the superconducting state?} 2) \textit{What is the origin of the $T_0$-anomaly observed at temperatures slightly above the superconducting state?} It appears that both these questions might be closely intertwined, and the first, most important step to get answers to them is to obtain single-crystalline samples of excellent quality. 
	
	Because our attempt to achieve prime-quality single crystals by repeating the previously reported procedure \cite{Khim2021} gave unsatisfactory results (more details in the text below), we have turned to a slightly different technique, namely flux growth in a horizontal configuration. Here we present our adaptation of this relatively rarely utilized but technically simple and in our opinion extremely powerful method for growing high-quality single crystals of CeRh$_2$As$_2$.
	
	\section*{Synthesis design}\label{sec2}
	
	As a starting point of our study and a reference for new crystal growth attempts, we decided to repeat the synthesis of CeRh$_2$As$_2$ single crystals following the previously reported routine (see supplementary material to Ref.~\cite{Khim2021} for details) and using a total charge of about 10~g. In this manner, we were able to obtain relatively small platelet-like single crystals with masses occasionally exceeding 1~mg. Their chemical composition, according to scanning electron microscopy with energy dispersive x-ray spectroscopy (SEM-EDS), is in good agreement with nominal stoichiometry 1:2:2, and their powder x-ray diffraction pattern matches with the tetragonal CeBe$_2$Ge$_2$-type crystal structure (not shown here). Despite this, the characteristic features of CeRh$_2$As$_2$ observed in the specific heat data, are distinctly smeared (see the discussion below), as was also reported before \cite{Khim2021,Hafner2022}. Most probably, it is due to the existence of some homogeneity range in the Ce-Rh-As system around 1:2:2 composition and the relatively large temperature interval (1150~--~700$^{\circ}$C) used during the synthesis process. In the conventional flux method, where crystal growth is mainly driven by temperature change and related shift of phase equilibrium, this may lead to crystallization of material with slightly divergent chemical composition, and therefore also with somewhat different properties, during subsequent stages of the growth. As a final product of such a process one obtains crystals possessing nonuniform bulk properties. Nevertheless, it should be noticed that also distinct temperature variations in the concentration of structural defects in the material grown may strongly influence its physical properties and lead to a similar inhomogeneity effect.
	
	In order to avoid such problems, we propose a different synthesis strategy based on the flux growth method in a horizontal configuration. In this technique, the starting material is sealed in an evacuated quartz tube with the constituent elements (in this case Ce, Rh, and As) placed at one end of the ampoule and flux material (here Bi) filling about 70-80 \% of the remaining volume. Such prepared ampoule is placed in a horizontal temperature gradient with substrate elements situated on the hot side. During the synthesis, several different zones, where various processes take place, can be distinguished  in the ampule. On the hot side (the dissolution zone) all constituent elements dissolve in the flux. Because there is a huge excess of these ingredients in comparison to their solubility limit in the flux metal and to the flux amount itself, only a small part of them is dissolved at a given moment. The remaining material serves as a reservoir of ingredients for further growth. In the intermediate section (the transport zone), dissolved elements are transported towards the other side of the ampule. The flux metal plays the role of transporting agent, and the concentration gradient is a driving force in this process. It is important to note that at the same time flux moves in the opposite direction allowing further dissolving of substrates in the hot zone. In the colder end of the ampule, due to lowering the solubility limit, desired material precipitates out of the flux and the formation of crystals takes place. Thus, this area can be called the crystallization (or growth) zone. 
	
	The most important feature of the horizontal flux growth method is its temperature stability. After some initial heating-up period, the temperature in each ampoule zone is kept stable throughout the whole synthesis process.
	Consequently, the crystals form in unaltered thermal conditions, which in turn leads to the crystallization of material showing highly uniform properties.
	Furthermore, because of the presence of opposite concentration gradients of the flux and the constituent elements, the feed material is continuously supplied to the crystallization zone. Hence, the growth process can be easily extended in time, making it possible to obtain crystals of large size. Theoretically, synthesis can be conducted until all substrates are dissolved in the flux and their concentration is equalized in the whole ampoule or until the transport channel becomes clogged by the previously grown crystals.
	
	\begin{figure}[htbp]%
		\centering
		\includegraphics[width=0.9\textwidth]{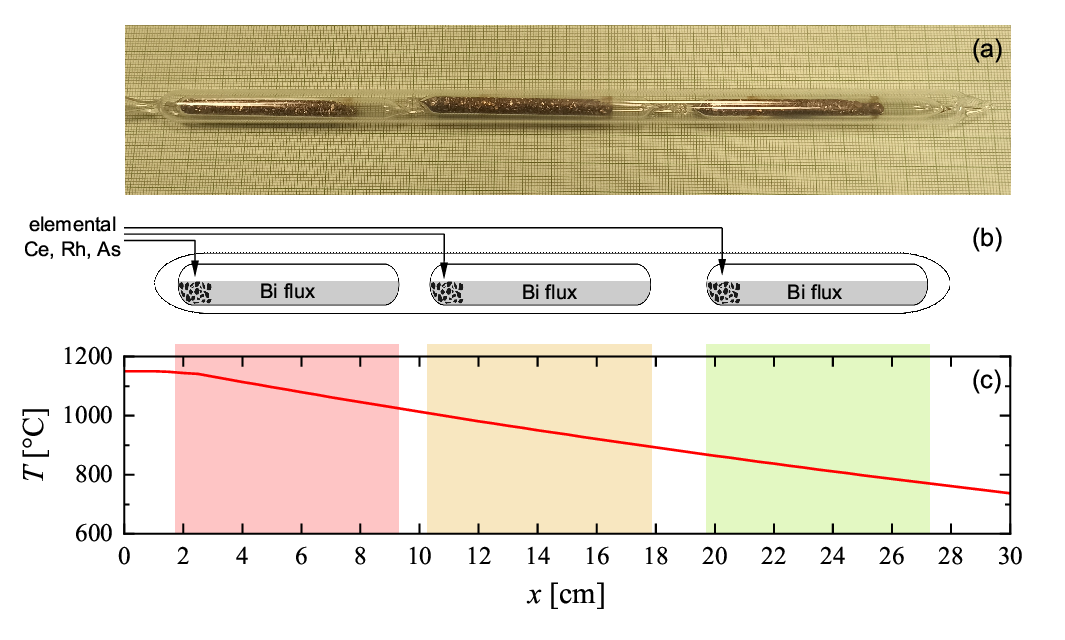}
		\caption{\textbf{Synthesis design.} \textbf{a} The photograph of a three-part test ampoule prepared for a horizontal flux growth; \textbf{b} Schematic picture of the ampoule with a marked arrangement of the elements inside; \textbf{c} temperature profile of the furnace; Three temperature sub-zones, where the ampoules were located during the growth, are marked by different colors.}\label{fig1}
	\end{figure}
	
	For our approach to the synthesis of CeRh$_2$As$_2$ single crystals via horizontal flux growth technique, we prepared three separate quartz tube ampoules of 6~mm inner diameter. All of them were filled in with total charges of about 8~g having the same initial composition as in the previous growth protocols, namely Ce:Rh:As:Bi~=~1:2:2:30. All the ingredients were placed in the ampoules in the manner described above and sealed in a vacuum. Such prepared ampoules were lined up in a row, enclosed in another, larger tube (see Fig.~\ref{fig1}a and Fig.~\ref{fig1}b),
	and located in the two-zone horizontal tube furnace with the temperature gradient inside set in such a way as to cover almost the entire temperature range used for the conventional flux growth. Fig.~\ref{fig1}c shows the temperature profile along the furnace.
	This way, we were able to distinguish three different temperature subranges: I - high-temperature range, II - medium-temperature range, and III - low-temperature range (red, yellow and green areas in Fig.~\ref{fig1}c, respectively), and to conduct in one furnace three separate syntheses at the same time.
	Such a procedure was aimed at determination of the most favorable temperature conditions for the formation of the CeRh$_2$As$_2$ crystals. The approximate temperature ranges in zones I, II, and III were 1150--1020 $^{\circ}$C, 1000--890 $^{\circ}$C, and 860--770 $^{\circ}$C, respectively.
	The tube was removed from the hot furnace after 15 days of growth.

	\section*{Results and Discussion}\label{sec3}
	
	\subsection*{Crystal growth}
	
	\begin{figure}[htbp]%
		\centering
		\includegraphics[width=0.9\textwidth]{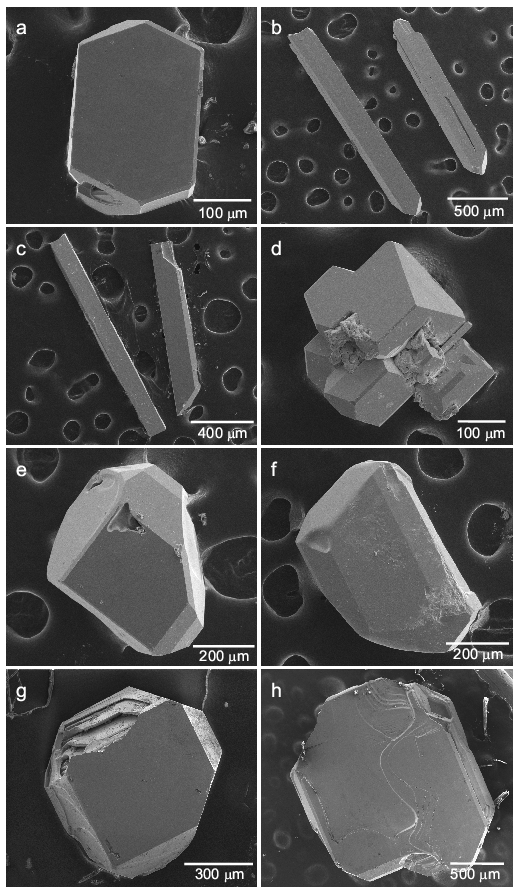}
		\caption{\textbf{Images of the grown single crystals.} Scanning electron microscope images of the representative single crystals obtained using horizontal flux growth method: \textbf{a}, \textbf{b} RhAs$_2$ crystals  grown in the high-temperature zone; \textbf{c}, \textbf{d} RhAs$_2$ and \textbf{e}, \textbf{f} CeRh$_2$As$_2$ crystals grown in the medium-temperature zone; \textbf{g} CeRh$_2$As$_2$ crystal grown in the low-temperature zone; \textbf{h} CeRh$_2$As$_2$ crystal obtained in the repeated synthesis (see the text for details).}\label{fig2}
	\end{figure}
	
	Fig.~\ref{fig2} shows scanning electron microscope images of some representative single crystals grown using the horizontal flux growth method within particular temperature ranges. In the ampoule from the high temperature zone we found crystals exhibiting two different morphologies, namely platelet- (Fig.~\ref{fig2}a) and blade-like (Fig.~\ref{fig2}b). According to the SEM-EDS results, both types had a chemical composition very close to RhAs$_2$. In this batch no crystals of CeRh$_2$As$_2$ were found. This suggests that above about 1000 $^{\circ}$C synthesis process is dominated by the crystallization of Rh-As binary compound and the desired ternary material forms at lower temperatures. 
	
	In the batch from the medium temperature zone, our findings were similar. Again, most of the collected crystals had blade- (see Fig.~\ref{fig2}c) or platelet-like shapes, but also some prismatic ones (Fig.~\ref{fig2}d) could be found. All of them are chemically composed only of Rh and As in a molar ratio close to 1:2. However, the crystals grown in the range II were slightly smaller than those synthesized at higher temperatures. Importantly, apart from the binary compounds, in this batch, we were also able to find small crystals with masses of about 0.5~mg and shapes suggesting tetragonal crystal structure (see Fig.~\ref{fig2}e and Fig.~\ref{fig2}f). Based on the results of SEM-EDS measurements they were confirmed to be crystals of CeRh$_2$As$_2$. As can be seen in Fig.~\ref{fig2}f, some of them have distinctly rounded edges, which may be the consequence of growth in the closeness to the melting temperature of the compound. 
	
	The best conditions for the growth of CeRh$_2$As$_2$ were found within the low-temperature range. In the latter batch, we obtained almost solely single crystals of the desired compound (except for very few small crystals of RhAs$_2$). They are visibly larger (see Fig.~\ref{fig2}g) than those grown at higher temperatures and the masses of the biggest ones are about 2.5~mg. Moreover, instead of common for tetragonal crystal structure platelet-like morphology, many of the crystals have comparable sizes in all three directions, which is very beneficial in the context of the characterization of their physical properties taking into account the crystal anisotropy.
	
	In order to verify the repeatability as well as scalability of the proposed synthesis method, we prepared an additional batch using a three times larger amount of materials (about 24~g of the total charge and the same composition as previously). For the growth, we set the same temperature at the hot end of the ampoule as it was used for the synthesis in the low-temperature range process  (860~$^{\circ}$C). Due to the use of a longer quartz tube, the temperature at the cold end of the ampoule was distinctly lower and was approximately 720~$^{\circ}$C. Furthermore, we extended the synthesis time up to 25 days. As a result, we obtained larger CeRh$_2$As$_2$ single crystals (see Fig.~\ref{fig2}h) with masses reaching 8~mg. Moreover, because after the synthesis, at the hot end of the ampoule, we found significant amount of starting materials undissolved in the flux, we expect that the process could be continued in time in order to achieve even more massive crystals. 
	
	\subsection*{Physical properties characterization}
	
	\begin{figure}[htbp]%
		\centering
		\includegraphics[width=0.88\textwidth]{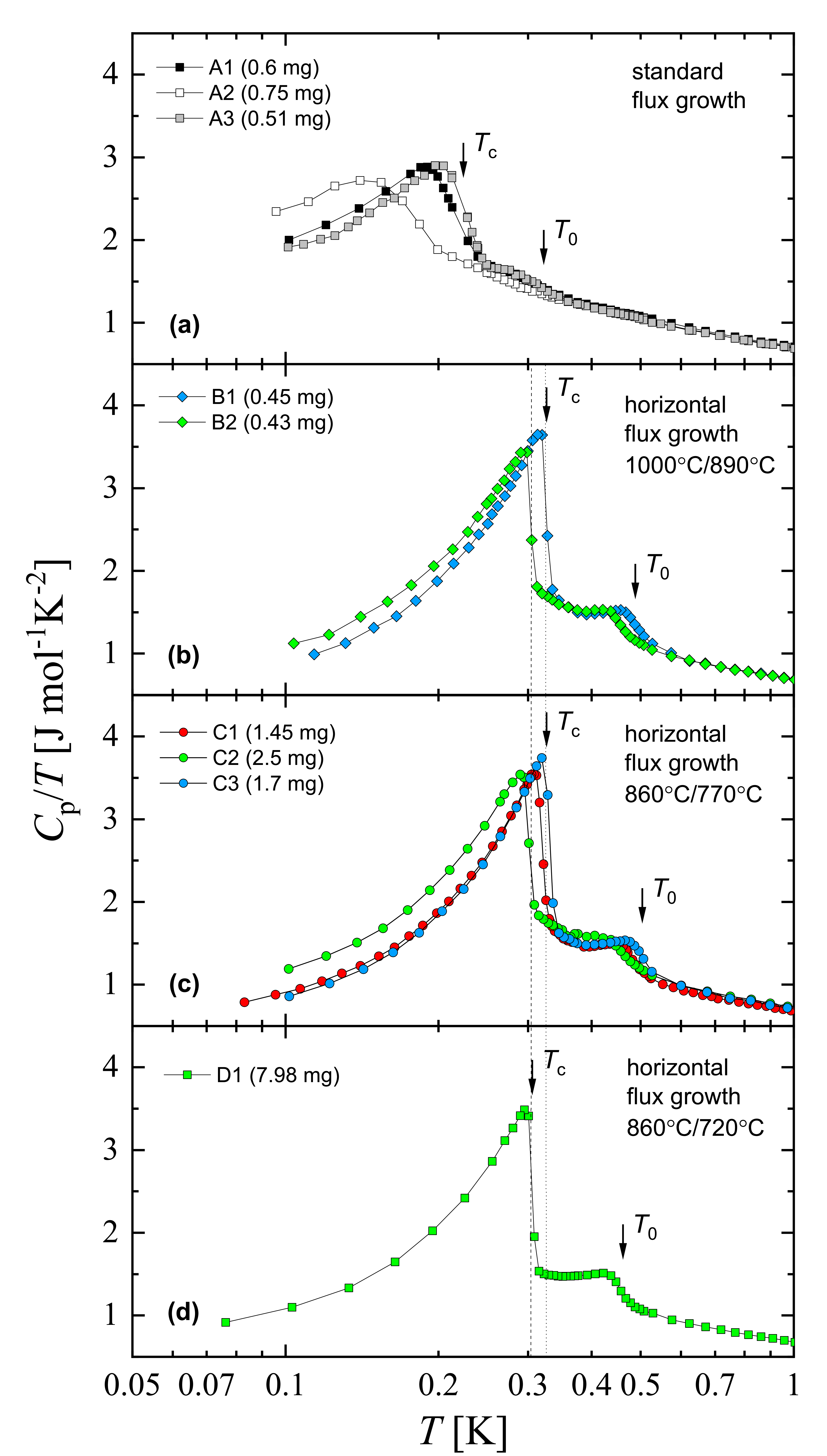}
		\caption{\textbf{Specific heat characterization of the crystals.} Low-temperature variations of the specific heat of the CeRh$_2$As$_2$ crystals grown by \textbf{a} standard flux method and \textbf{b--d} horizontal flux method in the medium-temperature zone, low-temperature zone, and during repeated growth, respectively. Arrows mark the phase transitions and vertical dashed and dotted lines are the guides for the eye. For the sake of clarity, in each panel, phase transitions were marked only for the curves with the highest transition temperatures.} \label{fig3}
	\end{figure}
	
	Due to the fact that in the previous reports on CeRh$_2$As$_2$ the crystal quality problem most clearly manifested itself through distinctly blurred temperature characteristics of various physical quantities, we decided to use heat capacity measurements as a reliable method to evaluate the quality of the obtained single crystals. The results of our experiments performed on samples from different batches are gathered in Fig.~\ref{fig3} and displayed as $C_{\rm p}/T (T)$ curves. 
	
	As can be easily inferred from Fig.~\ref{fig3}a, the standard flux growth method leads to the formation of crystals, which exhibit smeared specific heat anomalies related to the superconducting and 'the other' phase transitions at $T_{\rm c}$ and $T_0$, respectively. In addition, for some of the crystals (e.g. sample A2) the anomaly at $T_0$, rather than having a peak-like shape, appears to be only a change in the slope of the $C_{\rm p}/T (T)$ curve. Furthermore, in so-obtained crystals, the critical temperature $T_{\rm c}$ seems to be a sample-dependent property and varies significantly for different specimens. The highest values of the transition temperatures for this group of samples are $T_{\rm c}$~=~0.22~K and $T_{0}$~=~0.32~K.
	
	In contrast, all the CeRh$_2$As$_2$ crystals grown by the horizontal method exhibit a very sharp transition to the superconducting state at $T_{\rm c}$ and well defined specific heat anomaly at $T_0$ (see panels b--d in Fig.~\ref{fig3}). Furthermore, the magnitudes of both peaks in $C_{\rm p}$ are much larger and temperatures $T_{\rm c}$ and $T_0$ are distinctly higher, compared to those for samples grown by the standard flux technique. Interestingly, among numerous crystals tested, we were able to distinguish three groups of samples characterized by slightly different critical temperatures, namely $T_{\rm c}~=~$0.305(5), 0.318(6), and 0.330(8)~K (green, red, and blue points in Fig.~\ref{fig3}b--d, respectively). In order to maintain the clarity of the image, only one curve for each critical temperature observed in the batch is shown. One can also note, that for all of those crystals, samples with higher $T_{\rm c}$ have also higher $T_0$. This suggests that the two phase transitions in CeRh$_2$As$_2$ are linked to each other. Therefore, the related phenomena presumably interfere, and the unusual properties of the superconducting state in CeRh$_2$As$_2$ might result from the closeness of the phase transition at $T_0$.
	
	\begin{figure}[ht]%
		\centering
		\includegraphics[width=0.9\textwidth]{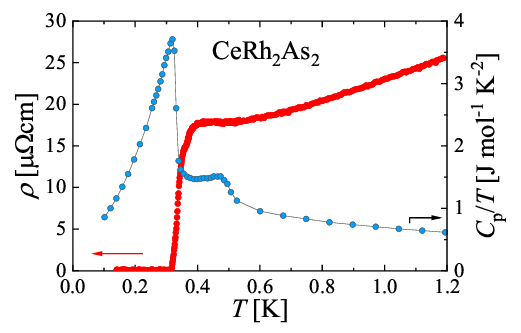}
		\caption{\textbf{Comparison of specific heat and electrical resistivity.} Low-temperature specific heat and electrical resistivity (current flowing within the tetragonal plane) of CeRh$_2$As$_2$ crystal grown in the low-temperature zone.}\label{fig4}
	\end{figure}
	
	Fig.~\ref{fig4} shows the comparison of the specific heat and electrical resistivity data collected on the same crystal grown in the low-temperature range process. As can be seen, in both measured quantities the superconducting phase transition occurs at almost ideally the same temperature.  This finding differs from the previous reports \cite{Khim2021, Hafner2022}, where the drops in the resistivity were observed at notably higher temperatures, compared to the anomalies in the specific heat and magnetic susceptibility, probably because of inhomogeneity in the chemical composition of the specimens. Thus, the result shown in Fig.\ref{fig4} can be considered as another proof of highly uniform properties of the crystals grown by the horizontal flux growth method.
	
	\subsection*{Crystal structure analysis}
	
	\begin{figure}[!ht]%
		\centering
		\includegraphics[width=0.95\textwidth]{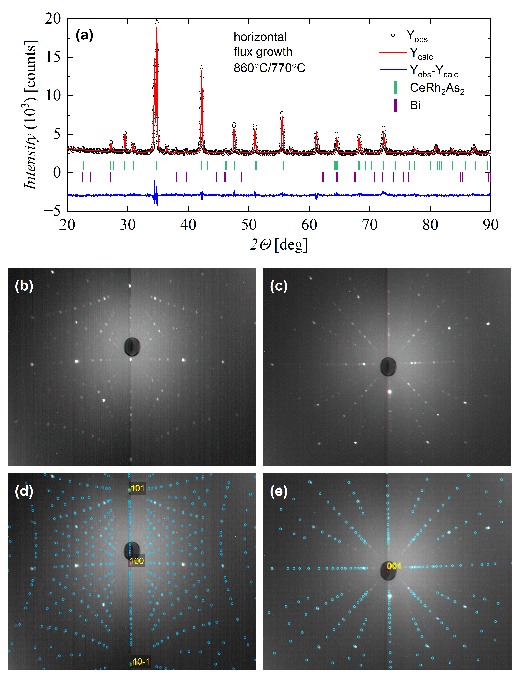}
		\caption {\textbf{Crystal structure analysis.} \textbf{a} Powder XRD pattern (circles) obtained for several pulverized single crystals of CeRh$_2$As$_2$ grown by the horizontal flux growth method in the low-temperature zone, together with the refined theoretical profile (red solid line), the difference between them (blue solid curve), and the refined positions of the Bragg reflections for the main phase and Bi flux residues (vertical ticks); \textbf{b} and \textbf{c} Laue x-ray backscattering patterns obtained for a single crystal of CeRh$_2$As$_2$ from the same batch for [100] and [001] directions, respectively; \textbf{d} and \textbf{e} the same images with the overlays of theoretically generated patterns (blue circles). } \label{fig5} 
	\end{figure}
	
	The powder x-ray diffraction (XRD) pattern collected on several ground single crystals from the low-temperature range batch is displayed in Fig.~\ref{fig5}a. As illustrated, all of the Bragg peaks can be ascribed to either CeRh$_2$As$_2$, crystallizing in the tetragonal CaBe$_2$Ge$_2$-type structure (space group \textit{P4/nmm}), or a small amount of residual Bi flux. The Rietveld analysis of this XRD data yielded the lattice parameters $a$ =4.2833(5)~\AA \ and $c$ = 9.8518(4)~\AA, in good agreement with the previously reported ones \cite{Khim2021}.
	
	The high quality of crystals grown by the horizontal flux  method was also confirmed by the Laue x-ray backscattering technique. The patterns collected in both [100] and [001] directions (figs.~\ref{fig5}b and \ref{fig5}c, respectively) show very sharp spots and no signatures of crystal twinning or other crystal defects. Moreover, they perfectly match with the theoretical patterns (figs.~\ref{fig5}d and \ref{fig5}e), generated based on the crystal structure parameters determined via the Rietveld refinement.
	
	\section*{Conclusion}\label{sec13}
	
	We presented the Bi flux growth technique in a horizontal configuration as an efficient method of obtaining large and high-quality single crystals of the unconventional superconductor CeRh$_2$As$_2$. The specific heat and electrical resistivity measurements, performed on numerous samples, proved that the synthesized single crystals exhibit highly uniform physical properties. Moreover, they show much sharper phase transitions, and located at higher temperatures, compared to the previously published data. The excellent quality of the grown crystals was also corroborated by the powder x-ray diffraction and Laue x-ray backscattering, and their chemical composition was confirmed to be very close to the nominal one by the SEM-EDS analysis.  Importantly, this growth technique was found to be easily scalable and highly repeatable.
	
	\section*{Methods}\label{sec4}
	\subsection*{Single crystal growth}
	Single crystals of CeRh$_2$As$_2$ were grown using two different approaches to the flux growth method: i) the conventional one - according to the procedure reported previously by Khim {\textit {et al.}} \cite{Khim2021} and ii) the horizontal flux growth method \cite{Yan2017} described in more detail in the main text. In the latter method, a horizontal two-zone tube furnace was used, however only in one of the zones (the hot zone) the heating was turned on during the whole synthesis process. In each of the batches the initial stoichiometry Ce:Rh:As:Bi~=~1:2:2:30 was used. In order to extract single crystals and remove flux, the material after synthesis was transferred to alumina crucibles and closed together with alumina strainers and additional reverted crucibles in evacuated quartz tubes. Such prepared ampules were reheated up to 450$^{\circ}$C and centrifuged. The residual flux remaining on the surfaces of crystals was removed by etching with the use of 2:1 mixture of glacial acetic acid and 30\% hydrogen peroxide. 
	
	\subsection*{X-ray diffraction and elemental analysis}
	The quality of the samples was checked using a PROTO COS Laue diffraction camera. The x-ray diffraction measurements were performed on several ground small single crystals on an X'pert Pro PANalytical diffractometer with Cu K$_{\alpha}$ radiation. The obtained powder x-ray diffraction data were analyzed with the use of the Rietveld refinement method implemented in FullProf software \cite{Carvajal1993} 
	
	Scanning electron microscope (SEM) images were taken using an FE-SEM FEI NovaNano SEM 230 equipped with an EDAX Genesis XM4 spectrometer. The same instrument was used to check the chemical composition of the samples via electron-microprobe analysis with energy-dispersive x-ray spectroscopy (EDS). SEM images and EDS measurements presented in the paper were acquired using the acceleration voltage of 30~kV.

	\subsection*{Specific heat and resistivity measurements}
	
	Specific heat measurements were performed in temperatures down to 80~mK using the standard time-relaxation method \cite{Hwang1998} implemented in a commercial Quantum Design PPMS (Physical Property Measurement System) platform equipped with a $^3$He-$^4$He dilution refrigerator. Electrical resistivity was measured employing the ac four-probe technique on bar-shaped samples with the use of the same experimental platform.

	\medskip\noindent 
	\textbf{Author Contributions} 
	G.C. synthesized the samples and performed their physical and crystallographic characterization. D.S. took SEM images of crytals and performed SEM-EDS based analysis. G.C. and D.K. discussed the results and prepared the manuscript. G.C., D.S. and D.K. revised and corrected the manuscript.
	
	\medskip\noindent 
	\textbf{Conflict of Interest}
	The authors declare no conflict of interest.
	
	\bibliography{bibliography}
	
	
\end{document}